\documentclass[%
 reprint,aps, prl,  
superscriptaddress,
 amsmath,amssymb,
]{revtex4-1}

\usepackage{graphicx}
\usepackage{dcolumn}
\usepackage{bm}
\usepackage[utf8]{inputenc}
\usepackage[T1]{fontenc}
\usepackage{lmodern}
\usepackage[english]{babel}
\usepackage{siunitx}
\usepackage{microtype}
\usepackage{textcomp}
\usepackage{amssymb}
\usepackage{xcolor} 

\DeclareSIUnit\molar{\mole\per\cubic\deci\metre}
\DeclareSIUnit\Molar{M}
\DeclareSIUnit\fps{fps}

\begin{document}

\title{The Role of Stickiness in the Rheology of Semiflexible Polymers}

\author{Tom Golde}
\affiliation{Peter Debye Institute for Soft Matter Physics, University of Leipzig, 04103 Leipzig, Germany}
\affiliation{Institute for Bioengineering of Catalonia, The Barcelona Institute for Science and Technology, 08028 Barcelona, Spain}
\author{Martin Glaser}
\affiliation{Peter Debye Institute for Soft Matter Physics, University of Leipzig, 04103 Leipzig, Germany}
\affiliation{Fraunhofer Institute for Cell Therapy and Immunology, 04103 Leipzig, Germany}
\author{Cary Tutmarc}
\affiliation{Peter Debye Institute for Soft Matter Physics, University of Leipzig, 04103 Leipzig, Germany}
\author{Iman Elbalasy}
\affiliation{Peter Debye Institute for Soft Matter Physics, University of Leipzig, 04103 Leipzig, Germany}
\author{Constantin Huster}
\affiliation{Institute for Theoretical Physics, University of Leipzig, 04103 Leipzig, Germany}
\author{Gaizka Busteros}
\affiliation{Peter Debye Institute for Soft Matter Physics, University of Leipzig, 04103 Leipzig, Germany}
\author{David M. Smith}
\affiliation{Fraunhofer Institute for Cell Therapy and Immunology, 04103 Leipzig, Germany}
\affiliation{Peter Debye Institute for Soft Matter Physics, University of Leipzig, 04103 Leipzig, Germany}
\author{Harald Herrmann}
\affiliation{Molecular Genetics, German Cancer Research Center, 69120 Heidelberg, Germany}
\affiliation{Department of Neuropathology, University Hospital Erlangen, 91054, Erlangen, Germany}
\author{Josef A. Käs}
\affiliation{Peter Debye Institute for Soft Matter Physics, University of Leipzig, 04103 Leipzig, Germany}
\author{Jörg Schnauß}
\affiliation{Peter Debye Institute for Soft Matter Physics, University of Leipzig, 04103 Leipzig, Germany}
\affiliation{Fraunhofer Institute for Cell Therapy and Immunology, 04103 Leipzig, Germany}

\date{\today}

\begin{abstract}
Semiflexible polymers form central structures in biological material.
Modeling approaches usually neglect influences of polymer-specific molecular features aiming to describe semiflexible polymers universally.
Here, we investigate the influence of molecular details on networks assembled from filamentous actin, intermediate filaments, and synthetic DNA nanotubes.
In contrast to prevalent theoretical assumptions, we find that bulk properties are affected by various inter-filament interactions.
We present evidence that these interactions can be merged into a single parameter in the frame of the glassy wormlike chain model.
The interpretation of this parameter as a polymer specific stickiness is consistent with observations from macro-rheological measurements and reptation behavior.
Our findings demonstrate that stickiness should generally not be ignored in semiflexible polymer models.
\end{abstract}

\maketitle

Semiflexible polymers play a central role in biological systems as major building blocks of intracellular scaffolds and extracellular matrices.
Among the most abundant semiflexible cytoskeletal biopolymers are filamentous actin (F-actin) and intermediate filaments (IF) \cite{huber_emergent_2013,herrmann_intermediate_2007}.
Network structures formed by these polymers exhibit unique viscoelastic properties, which cannot be easily deduced from the well-established theoretical frameworks for linear flexible polymers or rigid rods \cite{broedersz_modeling_2014}.
Classical polymer physics theories typically try to avoid details of molecular properties and mostly reduce semiflexible polymers to their size and stiffness in order to establish universal models.
Networks are modeled either as entangled networks, as in the tube model \cite{isambert_dynamics_1996,hinner_entanglement_1998,morse_viscoelasticity_1998-1}, or as cross-linked networks as in the affine model \cite{mackintosh_elasticity_1995}.
Many features of the tube model, such as the scaling of the plateau modulus with monomer concentration and the behavior of single filaments within a network, indeed fit very well to the experimental data for F-actin \cite{kas_f-actin_1996,isambert_dynamics_1996}.
The affine model has been demonstrated to predict the correct scaling of the plateau modulus in terms of concentration and cross-linker density for cross-linked F-actin \cite{gardel_elastic_2004,tharmann_viscoelasticity_2007}, but also for vimentin and keratin IF in the presence of $\text{MgCl}_2$ \cite{lin_origins_2010,leitner_properties_2012,pawelzyk_mechanics_2013}.

In reality, however, biopolymers without added cross-linkers already display adhesive interactions partially screened by electrostatic repulsion \cite{piazza_protein_2004,schopferer_desmin_2009,janmey_polyelectrolyte_2014,semerdzhiev_hydrophobic-interaction-induced_2018}.
For F-actin, minor impurities and aging effects are reported to cause a strong batch-to-batch variation of the network properties \cite{morse_viscoelasticity_1998-1,xu_mechanical_1998}.
IF networks feature pronounced hydrophobic interactions causing a weak concentration scaling of the network stiffness and a pronounced strain-stiffening in the non-linear deformation regime \cite{yamada_mechanical_2003,pawelzyk_attractive_2014,block_physical_2015}.
These effects can neither be explained by the tube nor by the affine deformation model.
Furthermore, recent experimental studies on F-actin and DNA-based semiflexible polymers present evidence that central predictions of the tube model in respect to the persistence length $l_\text{p}$ might be false \cite{schuldt_tuning_2016,tassieri_dynamics_2017}.


Here, we employ the natural filaments F-actin, vimentin and keratin IF as well as purely artificial double-crossover DNA nanotubes (DX tubes) in order to investigate the influence of (unspecific) adhesive interactions on the rheology of semiflexible polymer networks.
DX tubes are used as an additional synthetic semiflexible polymer model-system based on the self-assembly of DNA tiles \cite{rothemund_design_2004,ekani-nkodo_joining_2004}.
These tubes, with a diameter between $7$ and \SI{20}{\nano\metre}, a persistence length of around \SI{4}{\micro\metre}, a contour length of several micrometers, and a negative surface charge, were chosen for their similarity to the biopolymers under investigation (Table SI \cite{Supplement}).

\begin{figure}[tb]
	\centering
		\includegraphics{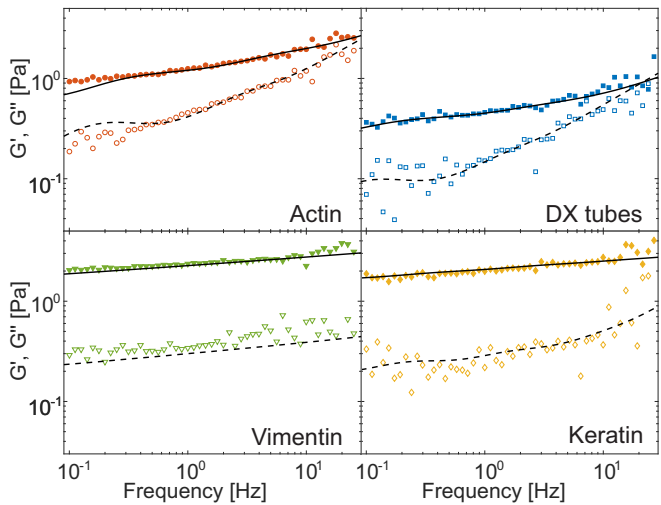}
	\caption{Typical storage modulus $G'$ (solid symbols) and loss modulus $G''$ (open symbols) versus frequency. Black lines are the result of fitting G'(solid) and G''(dashed) simultaneously with Eq.(\ref{eq:GWLC_G}) of the GWLC. Although the curves roughly resemble a rubber plateau, they in fact follow a weak power law. The cross-over frequency between $G'$ and $G''$ significantly varies for different polymer types. 
}
	\label{fig:Figure1}
\end{figure}

FIG. \ref{fig:Figure1} displays typical results for F-actin (\SI{0.5}{\gram\per\litre}), DX tubes (\SI{1}{\gram\per\litre}), vimentin (\SI{1}{\gram\per\litre}) and keratin K8/K18 (\SI{0.5}{\gram\per\litre}) IF networks.
The monomer concentrations were chosen in order to have a comparable mesh size (See Supplemental Material \cite{Supplement}).
All networks feature a storage modulus $G'$ that appears flat, but in fact behaves like a weak power law.
F-actin and DX tubes display a beginning cross-over between $G'$ and the loss modulus $G''$ while the cross-over frequency for vimentin and keratin IF has been previously shown to appear at frequencies higher than probed by macro-rheology \cite{pawelzyk_attractive_2014}.
The cross-over of $G'$ and $G''$ denotes the transition from network properties dominated by filament interactions for low frequencies to the high frequency regime dominated by single filament behavior \cite{gittes_dynamic_1998}.

In order to enable a quantitative comparison of different polymers and samples, we characterized the shear modulus via an approximation of the elastic modulus by a local power law $G'(\omega) \propto \omega^\alpha$ with exponent $\alpha$ in the frequency regime below the cross-over.
For IF, the local power law exponent $\alpha$ is around 0.07 and it is about twice as large for F-actin and DX tubes with $\alpha \approx 0.14$ [FIG. S1 \cite{Supplement}].
Additionally, we display the loss factor $\textrm{tan}(\phi) = G'' / G'$ at a fixed frequency of \SI{1}{\hertz}.
This frequency was chosen to avoid experimental noise in the low frequency regime while still maintaining a frequency independent loss factor for most samples [FIG. S2 \cite{Supplement}].
The loss factor has a strong sample to sample variation for F-actin ($\textrm{tan}(\phi)=0.40 \pm 0.11$) and decreases over DX tubes and vimentin to keratin ($\textrm{tan}(\phi) =0.11 \pm 0.02$), meaning the networks become more elastic [FIG. S1 \cite{Supplement}].

We recently demonstrated that $\alpha$ and $\textrm{tan}(\phi)$ of composite networks of actin and vimentin filaments have intermediate values in comparison to pure networks \cite{golde_composite_2018}.
They can be tuned simply by the ratio of actin and vimentin filaments in the network.
However, both the tube and the affine model predict only a flat plateau for frequencies below the cross-over \cite{mackintosh_elasticity_1995,isambert_dynamics_1996,morse_viscoelasticity_1998-1}.
These models are typically used to compare only the scaling predictions of the network stiffness with experimental data \cite{morse_viscoelasticity_1998-1,hinner_entanglement_1998,mackintosh_elasticity_1995,isambert_dynamics_1996,Kroy_Force_1996,morse_tube_2001,gardel_microrheology_2003,gardel_elastic_2004,liu_microrheology_2006,tharmann_viscoelasticity_2007,tassieri_dynamics_2008,atakhorrami_scale-dependent_2014,schuldt_tuning_2016,tassieri_dynamics_2017}.
To our knowledge, a study by Schmidt et al. presents the sole fit of the tube model to a measured frequency dependence of $G'$ and $G''$ and shows only a rough agreement \cite{schmidt_viscoelastic_2000}.
Thus, we need a different model for explaining the actual frequency dependence of the complex shear modulus.

The observed weak power laws are reminiscent of soft glassy systems and have been shown to be a main feature of micro-rheological experiments in cells, as well \cite{fabry_scaling_2001}.
A phenomenological model providing a description of the weak power law behavior on a network level is the glassy wormlike chain model (GWLC) established by Kroy and Glaser \cite{kroy_glassy_2007}.
This model is an extension of the wormlike chain (WLC), the minimal model of a semiflexible polymer.
The constituting idea is that the mode relaxation times $\tau_{n}$ of all eigenmodes of (half-)wavelength $\lambda$ and modenumber $n$ that are longer than a characteristic interaction length $\Lambda$ are stretched exponentially:
\begin{equation}\label{eq:GWLC_tau}
\tau_{n}^{\text{GWLC}} = 
\begin{cases} 
\tau_{n}^{\text{WLC}} \hspace{0.7cm}   \hspace{0.7cm}  &\mbox{ if } \lambda_{n} \le \Lambda \\ 
\tau_{n}^{\text{WLC}} \text{e}^{\varepsilon N_n}         &\mbox{ if } \lambda_{n} > \Lambda.  
\end{cases}
\end{equation}
Here, $N_n = \lambda_n / \Lambda - 1$ is the number of interactions per length $\lambda$.
$\varepsilon$ is the stretching parameter controlling how strong the modes are slowed down.
The assumption of an exponential stretching is directly supported by the experimental observation of logarithmic tails of the dynamic structure factor in F-actin solutions \cite{semmrich_glass_2007}.
The complex linear shear modulus in the high frequency regime is: 
\begin{equation}\label{eq:GWLC_G}
G^{*}(\omega) = \Lambda/(5 \xi^2 \chi(\omega)),
\end{equation}
where $\xi$ is the mesh size and $\chi(\omega)$ is the micro-rheological, linear response function to a point force at the ends of the GWLC at frequency $\omega$. 
The specific model used for this study has been comprehensively described previously \cite{golde_composite_2018} and more details are presented in the Supplemental Material \cite{Supplement}.

We can fit Eq. (\ref{eq:GWLC_G}) directly to the macro-rheological data in the linear regime for each sample.
The fit parameters are the mesh size $\xi$, the interaction length $\Lambda$ and the stretching parameter $\varepsilon$.
All other parameters were fixed to literature values or experimentally obtained (see Table SI \cite{Supplement}).
$\Lambda$ is determined by the cross-over frequency $\omega_{\Lambda} = {2 \pi^5 l_\text{p} {k_{\text{B}}T}/\left( {\zeta_{\perp} \Lambda^4}\right)}$ with transverse drag coefficient $\zeta_{\perp}$.
For the fitting of the IF networks, $\Lambda$ is assumed to be below, but of the same order of magnitude as $\Lambda$ for F-actin and DX tubes to account for a larger $\omega_{\Lambda}$ [FIG. S7 \cite{Supplement}].
$\varepsilon$ is the parameter that defines the functional dependence of $G^*(\omega)$ for frequencies $\omega < \omega_\Lambda$.
$\xi$ is used as a free fit parameter to obtain the correct network stiffness because it only shifts the magnitude of $G^*$ without any influence on the functional dependency.

\begin{figure}[t]
	\centering
		\includegraphics{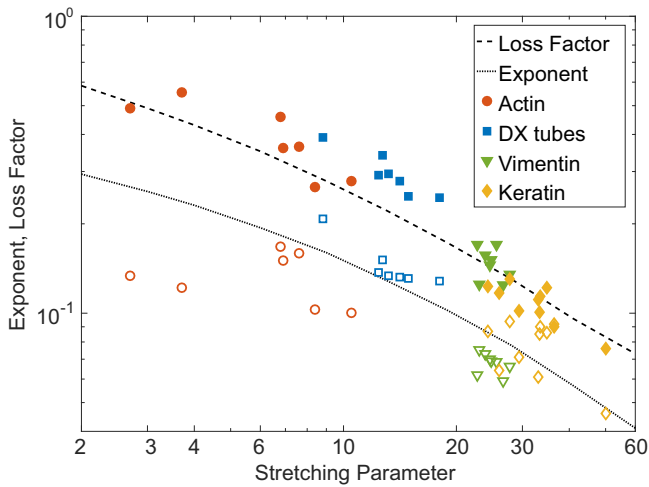}
	\caption{Local power law exponent of $G' \propto \omega^\alpha$ (open symbols) and loss factor $\textrm{tan}(\phi)=G''/G'$ (solid symbols) versus stretching parameter $\varepsilon$. Each pair of data points represents one sample. The exponent was obtained from fitting $G'$ with a power law for frequencies smaller than the cross-over between $G'$ and $G''$. The loss factor was obtained from fitting $\textrm{tan}(\phi)$ locally with a power law at a frequency of \SI{1}{\hertz}. $\varepsilon$ is the result from fitting the complex shear modulus $G^*$ to Eq.(\ref{eq:GWLC_G}) for each sample. Dashed lines are the numerical results of an exemplary $G^*_\text{GWLC}$ where all parameters except $\varepsilon$ are fixed.}
	\label{fig:Figure2}
\end{figure}

We then compare $\varepsilon$ with the local power law exponent $\alpha$ and the loss factor $\tan(\phi)$ (FIG. \ref{fig:Figure2}).
It is worth noting that $\alpha$, $\tan(\phi)$ and $\varepsilon$ are directly connected in the theory and their relation can be approximated analytically for $\varepsilon \gg 1$ as presented in Ref. \cite{kroy_rheological_2009}.
The comparison reveals that a small loss factor correlates with a small power law exponent.
Moreover, we find significant differences in $\varepsilon$  between all polymer types with mean values $\pm$ standard deviation of $6.7 \pm 2.7$ for actin, $13.4 \pm 2.8$ for DX tubes, $24.9 \pm 1.7$ for vimentin and $31.8 \pm 7.2$ for K8/K18 [FIG. S1 \cite{Supplement}].

The original study by Kroy and Glaser suggests the interpretation of $\varepsilon$ as a kinetic "stickiness" parameter \cite{kroy_glassy_2007}.
$\varepsilon$ can be thought of as the height of the free energy barriers of unspecific filament-to-filament interactions in units of $k_{\text{B}}T$ \cite{semmrich_glass_2007}.
Later interpretations suggest a test polymer that can bind and unbind to "sticky" entanglement points by overcoming the energy barrier, which slows the contributions from long-wavelength bending modes during relaxation \cite{wolff_inelastic_2010}.
In the following, we will demonstrate that $\varepsilon$ appears indeed as a polymer specific stickiness that combines all filament-to-filament interactions into one number.

A look at the molecular details of biopolymers suggests several adhesive interactions as plausible candidates for sticky interactions.
Since semiflexible biopolymers are relatively massive multi-molecular assemblies, errors such as misfoldings or hydrophobic loops \cite{feng_fluorescence_1997} are expected to occur in general on a purely stochastic basis.
While F-actin networks are considered as a model system for entangled semiflexible polymer networks, the batch-to-batch variation is suggested to be caused by very small amounts of cross-links\cite{morse_viscoelasticity_1998-1,xu_mechanical_1998}.
DX tubes appear similar to F-actin in regards to their polyelectrolyte properties \cite{janmey_polyelectrolyte_2014}.
The higher $\varepsilon$ could be a consequence of mishybridiazation during the assembly.
The large $\varepsilon$ of IFs can be explained by their dominant hydrophobic interactions \cite{yamada_mechanical_2003,pawelzyk_attractive_2014}.
These interactions are partially mitigatied by electrotstactic repulsion between vimentin IF \cite{schopferer_desmin_2009}, leading to a smaller $\varepsilon$ in comparison to keratin.
A more detailed discussion of filament-to-filament interactions can be found in the Supplemental Material \cite{Supplement}.

The experimental data can be further compared to the model by calculating an exemplary $G^*_\text{GWLC}$.
The model parameters are fixed to contour length $L=\SI{18}{\micro\metre}$, $l_\text{p} = \SI{4}{\micro\metre}$, $\Lambda = \SI{1}{\micro\metre}$, and $\xi = \SI{0.2}{\micro\metre}$, resembling intermediate values of the experimental results, and only $\varepsilon$ is varied.
The obtained $G^*_\text{GWLC}(\varepsilon)$ are analyzed for each $\varepsilon$ in the same way as the rheological data.

Remarkably, the resulting curves for $\alpha_\text{GWLC}(\varepsilon)$ and $\tan(\phi)_\text{GWLC}(\varepsilon)$ can already be viewed as a master curve for the experimental data without any rescaling, although $L$,$l_p$,$\Lambda$ and $\xi$ differ for every polymer type [FIG. \ref{fig:Figure2}].
A better agreement between experimental data and the GWLC can be reached by calculating $G^*_\text{GWLC}(\varepsilon)$ for each polymer [FIG. S3 \cite{Supplement}].

In contrast to $\alpha$ and $\tan(\phi)$, $\varepsilon$ is significantly different for all four polymers [FIG. S1 \cite{Supplement}]. 
Thus, $\varepsilon$ is not only the key parameter of the GWLC, but it might be a key factor for deriving a universal master curve that unifies systems with fundamentally different molecular details, as well.
It appears as a very robust quantity for characterizing the linear rheological behavior of semiflexible polymer networks and seems to provide a more universal description of network properties than the pseudo plateau modulus $G_0$.

\begin{figure}[b]
	\centering
		\includegraphics{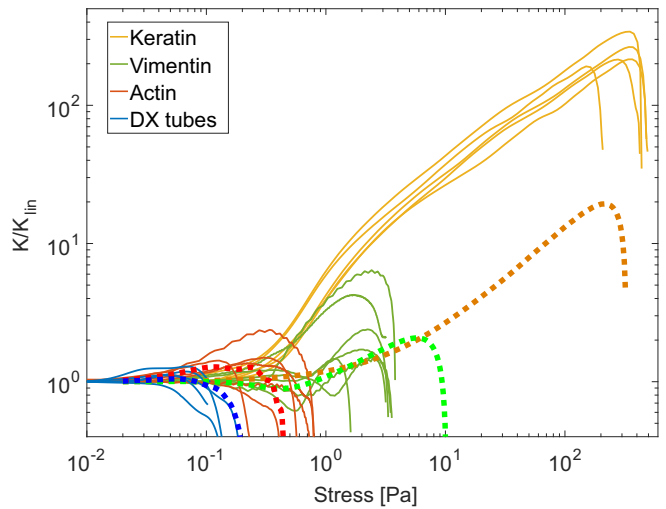}
	\caption{Differential shear modulus $K=\text{d}\sigma/\text{d}\gamma$ rescaled by its value in the linear regime $K_{\text{lin}}$ versus stress $\sigma$. Solid lines are single measurements. Dotted lines are replicated curves with the non-linear extension of the GWLC. F-actin and DX tubes have a similar behavior with weak to no strain-stiffening. Strain-stiffening is more pronounced for vimentin IF while keratin IF have the highest peak value $K_{\textrm{max}}$ at a much larger $\sigma$. F-actin and vimentin IF data reproduced from \cite{golde_composite_2018}.}
	\label{fig:Figure3}
\end{figure}

In contrast to the linear regime, it is well-known that the behavior of F-actin and IF at large deformations is drastically different to each other \cite{janmey_viscoelastic_1991,storm_nonlinear_2005}.
To investigate these differences, the GWLC can be extended to the non-linear regime and the differential shear modulus $K=\text{d}\sigma/\text{d}\gamma$, defined as the derivative of stress $\sigma$ over strain $\gamma$, can be measured with a $\dot{\gamma}$-protocol as described previously \cite{golde_composite_2018}(see Supplemental Material for details \cite{Supplement}).
Although $K$ is measured in dependency of $\gamma$ as in FIG. S4 \cite{Supplement}, it can be displayed over $\sigma$ to enable a comparison with the model [FIG. \ref{fig:Figure3}].

With this method, we are able to replicate $K$ for F-actin, DX tubes and vimentin filament networks [FIG. \ref{fig:Figure3}, Table SI \cite{Supplement}].
The initial softening of vimentin IF can be captured by an additional bond-breaking mechanism as described previously \cite{golde_composite_2018}.
For keratin K8/K18, we can shift the peak of $K$ to the correct $\sigma$, but underestimate $K_{\textrm{max}}$ by an order of magnitude.
The phenomenology of keratin IF is potentially based on strong filament interactions as well as a small $l_p$ leading to two different slopes for $K$ due to a cross-over from a bending to a stretching dominated regime.
This behavior is better described by a triangular lattice model for physiological cross-linked networks \cite{p.broedersz_molecular_2011}.
Keratin IF act as some kind of limiting case for the applicability of the GWLC in the non-linear regime.
The observation that the polymer with the largest $\varepsilon$ behaves more like a cross-linked network supports the interpretation of $\varepsilon$ as stickiness.


The constituting idea of the GWLC in Eq. (\ref{eq:GWLC_tau}) can be implemented using an effective mode dependent friction transversal to the filament $\zeta_n = \zeta_\perp \textrm{exp}(N_n\varepsilon)$ for $\lambda_n > \Lambda$ as well.
If this increase of the transversal friction is indeed caused by sticky interactions, this should also increase the longitudinal friction and slow down the reptation of single filaments within the network.
To test this reasoning, we observed embedded fluorescent tracer filaments and analyzed the mean-squared displacement (MSD) of the filament center parallel to the tangent vector as described previously \cite{schuldt_tuning_2016,golde_composite_2018}.
Unfortunately, this technique is not applicable for keratin because there is no live stain for native keratin IF leading to their exclusion from the following analysis.
For the following examination, the "tube" is simply the space formed by the geometrical constraints due to the surrounding filaments that can be probed by a test polymer \cite{doi_theory_1988}.

\begin{figure}[tb]
	\centering
		\includegraphics{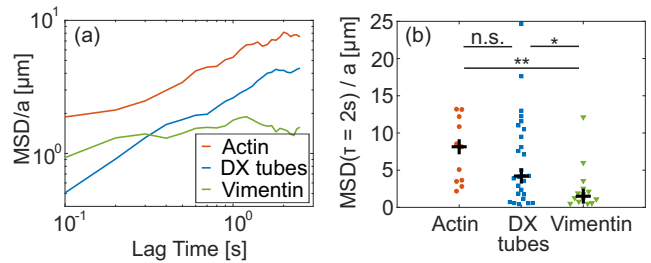}
	\caption{(a) MSD of the filament center parallel to the tube rescaled by tube width $a$ versus lag time $\tau$. The lines are the median of all observed filaments with $n\geq10$. (b) Distribution of MSD at lag time $\tau = \SI{2}{\second}$ rescaled by tube width $a$. Each data point is a single filament. The black cross denotes the median and illustrates that the overall motility decreases from F-actin over DX tubes to vimentin IF.}
	\label{fig:Figure4}
\end{figure}

A quantitative comparison between different polymer networks can be achieved by looking at the MSD at $\tau = \SI{2}{\second}$, where the MSD is in a weak power law regime, rescaled by the tube width $a$ [FIG. \ref{fig:Figure4}] \cite{granek_semi-flexible_1997,mcleish_tube_2002}.
In this regime, the MSD is independent of the polymer length.
We use the MSD instead of the more common longitudinal diffusion coefficient because there is no diffusion for the stick-slip like systems investigated here.

Both F-actin and DX tubes reveal a strong filament-to-filament variation with similar distributions.
The distribution of vimentin IF is dominated by significantly smaller values in comparison to both F-actin ($p=\num{1.6e-3}$) and DX tubes ($p=\num{3.1e-2}$) [FIG. 4(b), FIG. S5 \cite{Supplement}].
The main difference between F-actin and DX tubes is that some DX tubes have a flat MSD.
This implies that they are stuck at their respective position and hints at mishybridization during the assembly process.
Such a behavior was not observed for F-actin [FIG. S5, \cite{Supplement}].

In the tube model, the MSD$(\tau = \SI{2}{\second})/a$ is expected to increase for smaller persistence lengths because the mode of transportation is dominated by filament undulations in this time regime \cite{granek_semi-flexible_1997}.
Here, we see the exact opposite behavior.
This means the assumption behind the persistence length scaling is either violated or overwritten by an additional factor like the proposed effective friction.

Recent Brownian dynamics simulations of entangled solutions of semiflexible polymers by Lang and Frey demonstrate, that polymer relaxation might have to be considered as a many-body effect with dynamic correlations instead of a diffusive motion along a tube\cite{lang_disentangling_2018}.
Their simulations demonstrate that varying friction coefficients strongly influence the interplay of a tracer polymer with its surrounding .
This friction is not necessarily the same as the proposed sticky interactions.
Including stickiness, however, could provide further insight into relaxation processes within a network that seem to be more complicated than assumed by the tube model.

The motion of the tracer filaments could potentially be influenced by the attached fluorescent dye.
However, the labeling alone does not seem to impede filament motion in networks \cite{kas_f-actin_1996,schuldt_tuning_2016,keshavarz_confining_2017}.
Thus, the decrease of the MSD$(\tau = \SI{2}{\second})/a$ from F-actin over DX tubes to vimentin IF while $\varepsilon$ increases, supports the interpretation of $\varepsilon$ as a polymer stickiness, which increases the longitudinal friction and slows down filament motion.


Considering the discussed limitations, it is remarkable that the GWLC captures most of the linear and non-linear macro-rheological properties of the semiflexible polymer networks investigated here.
The stretching parameter $\varepsilon$ is more than a simple free fit parameter.
Our results consistently support the interpretation of $\varepsilon$ as a polymer specific stickiness that strongly affects rheological characteristics and might be able to overwrite scaling predictions in classical semiflexible polymer theories.
The different magnitudes of stickiness for F-actin and IF may help to get a better understanding of their roles in living cells.
Cells are able to modify network structures with numerous binding proteins, especially for F-actin.
However, inherent sticky interactions such as hydrophobic interactions for IF, would limit the ability to further tune network properties.
At the same time, the high stickiness of IF contributes to their non-linear behavior and possibly influences cell properties under large deformations.
We expect that the GWLC can also be used to analyze other sticky semiflexible polymers such as the recently investigated temperature dependent hydrophobic interactions in $\alpha$-synuclein fibril networks \cite{semerdzhiev_hydrophobic-interaction-induced_2018}.
While the simplistic phenomenological nature of the GWLC diminishes some explanatory power, it shows that inter-filament stickiness impacts semiflexible polymer networks and should be considered in polymer models aiming to fully describe the dynamics of such systems.
The large size and high complexity of semiflexible polymers makes filament misfolding and impurities likely, even for supposedly interactionless proteins like F-actin.
Including stickiness as a universal feature in semiflexible polymers means a paradigm shift in classical polymer physics because it allows to unify systems that where to date treated either as purely entangled or chemically cross-linked.
This approach might help to further explain and resolve the current discrepancies between established models and experimental data \cite{schuldt_tuning_2016,tassieri_dynamics_2017}.

\begin{acknowledgments}
We gratefully thank Tatjana Wedig (DKFZ) for technical assistance with vimentin and keratin procedures, and Klaus Kroy for fruitful discussions.
Furthermore, we acknowledge funding by the Deutsche Forschungsgemeinschaft for M.G.(DFG-1116/14-1) and H.H.(HE 1853/11-1) and by the European Research Council (ERC-741350). 
\end{acknowledgments}

	

%

\clearpage

\onecolumngrid
\begin{center}
  \textbf{\large Supplemental Material: The Role of Stickiness in the Rheology of Semiflexible Polymers}\\[.2cm]
  Tom Golde,$^{1,2}$ Martin Glaser,$^{1,3}$ Cary Turmarc,$^{1}$ Iman Elbalasy,$^{1}$ Constantin Huster,$^{4}$ Gaizka Busteros,$^{1}$ David M. Smith, $^{3,1}$ Harald Herrmann,$^{5,6}$ Josef A. Käs,$^{1}$ and Jörg Schnauß$^{1,3,*}$\\[.1cm]
  {\itshape ${}^1$Peter Debye Institute for Soft Matter Physics,
University of Leipzig, 04103 Leipzig, Germany\\
	${}^2$Institute for Bioengineering of Catalonia, The Barcelona Institute for Science and Technology, 08028 Barcelona, Spain\\
  ${}^3$Fraunhofer Institute for Cell Therapy and Immunology, 04103 Leipzig, Germany\\
  ${}^4$Institute for Theoretical Physics, University of Leipzig, 04103 Leipzig, Germany\\
	${}^5$Molecular Genetics, German Cancer Research Center, 69120 Heidelberg, Germany\\
	${}^6$ Department of Neuropathology, University Hospital Erlangen, 91054, Erlangen, Germany\\
  ${}^*$Electronic address: joerg.schnauss@uni-leipzig.de \\}
(Dated: \today)\\[1cm]
\end{center}

\setcounter{equation}{0}
\setcounter{figure}{0}
\setcounter{table}{0}
\setcounter{page}{1}
\renewcommand{\theequation}{S\arabic{equation}}
\renewcommand{\thefigure}{S\arabic{figure}}
\renewcommand{\thetable}{S\Roman{table}}

\section*{Materials and Methods}

\subsection{Keratin}
Recombinant human keratins K8 and K18 were expressed, purified and prepared as described in \cite{herrmann_structure_1996,herrmann_characterization_2002}.
Briefly, proteins were expressed in E. coli, purified and stored in \SI{8}{\Molar} urea at \SI{-80}{\celsius}.
Before use, K8 and K18 were mixed in equimolar ratios and renatured by dialysis against \SI{8}{\Molar} urea, \SI{2}{\milli\Molar} Tris–HCl (pH 9.0) and \SI{1}{\milli\Molar} DTT with stepwise reduction of the urea concentration (\SI{6}{\Molar}, \SI{4}{\Molar}, \SI{2}{\Molar}, \SI{1}{\Molar}, \SI{0}{\Molar}).
Each dialysis step was done for \SI{20}{\minute} at room temperature, then the dialysis was continued overnight against \SI{2}{\milli\Molar} Tris-HCl, pH 9.0, \SI{1}{\milli\Molar} DTT at \SI{4}{\celsius}.
The dialyzed protein was kept on ice for a maximum of four days.
The final protein concentration was determined by measuring the absorption at \SI{280}{\nano\metre} using a DU 530 UV/Vis Spectrophotometer (Beckman Coulter Inc., USA).
Assembly of keratin was initiated by addition of an equal volume of \SI{18}{\milli\Molar} Tris–HCl buffer (pH 7.0) to renatured keratins resulting in a final buffer condition of \SI{10}{\milli\Molar} Tris–HCl (pH 7.4). 

\subsection{Double-Crossover DNA Nanotubes}
All oligomers for hybridization of the DNA nanotubes were adapted from Ekani-Nkodo \textit{et al.} \cite{ekani-nkodo_joining_2004}[Table SII] and purchased from Biomers.net with HPLC purification.
In order to assemble a nanotube network of a desired concentration the required strands (SE1-SE5) were mixed in equimolar concentration in an assembly buffer containing \SI{40}{\milli\Molar} Tris-acetate, \SI{1}{\milli\Molar} EDTA and \SI{12.5}{\milli\Molar} $\text{Mg}^{2+}$ (pH 8.3).
The concentration of each stock solution was confirmed spectrophotometrically by a NanoDrop 1000 (Thermo Fisher Scientifc Inc., USA) at a wavelength of \SI{260}{\nano\metre}.
These strands were hybridized in a TProfessional Standard PCR Thermocycler (Core Life Sciences Inc.,USA) by denaturation for \SI{10}{\minute} at \SI{90}{\celsius} and complementary base pairing for \SI{20}{\hour} between \SI{80}{\celsius} and \SI{20}{\celsius} by lowering the temperature by \SI{0.5}{\kelvin} every \SI{10}{\minute}.
After hybridization DNA nanotubes were stored at room temperature.
For visualization the oligomer SE3 was modified with the fluorescent Cyanine dye 3 with two additional spacer thymine bases in between.
DNA nanotubes were labeled by partially or fully replacing the unlabeled oligo SE3 by SE3-Cy3.

\subsection{Actin}

Monomeric actin (G-actin) was obtained with an acetone powder prep from rabitt muscle, purified, and stored at \SI{-80}{\celsius} in G-Buffer (\SI{2}{\milli\Molar} sodium phosphate buffer pH 7.5, \SI{0.2}{\milli\Molar} ATP, \SI{0.1}{\milli\Molar} $\textrm{CaCl}_2$, \SI{1}{\milli\Molar} DTT, \SI{0.01}{\percent} $\textrm{NaN}_3$) as described previously \cite{gentry_buckling-induced_2009}.
Small sample volumes were thawed and kept on ice no longer than one day before experiments.
The polymerization to F-actin was always induced by adding $1/10$ volume fraction of 10 times concentrated F-Buffer (\SI{20}{\milli\Molar} sodium phosphate buffer pH 7.5, \SI{1}{\Molar} KCl, \SI{10}{\milli\Molar} $\textrm{MgCl}_2$, \SI{2}{\milli\Molar} ATP, \SI{10}{\milli\Molar} DTT) to the final sample solution.
F-actin was fluorescently labeled by polymerizing G-actin at \SI{5}{\micro\Molar} in a $1:1$ ratio with Phalloidin–Tetramethylrhodamine B isothiocyanate (Phalloidin-TRITC - Sigma-Aldrich Co., USA).

\subsection{Vimentin}

Human vimentin was obtained from recombinant expression in E.coli and purified from inclusion bodies as described by Herrmann \textit{et al.} \cite{herrmann_isolation_2004}.
Before the assembly into filaments, the purified vimentin was dialyzed in a stepwise fashion from \SI{8}{\Molar} urea against a \SI{2}{\milli\Molar} sodium phosphate buffer at pH 7.5 and kept on ice for a maximum of four days \cite{mucke_molecular_2004}.
Polymerization was induced as described for actin.
Fluorescent labeling was performed with  Alexa Fluor 488 C5 Maleimide (Thermo Fisher Scientifc Inc., USA) as described by Winheim \textit{et al.} \cite{winheim_deconstructing_2011}.
The only modification was the removal of excess dye by elution over PD-10 Desalting Columns (GE Healthcare, USA).
Unlabeled vimentin monomers were mixed with about \SI{10}{\percent} labeled monomers before dialysis to obtain fluorescently labeled filaments.

\subsection{Shear Rheology}

Shear rheology measurements were performed with a strain controlled ARES rheometer (TA Instruments, USA) equipped with a \SI{40}{\milli\metre} plate-plate geometry at a gap width of \SI{140}{\micro\metre}.
Biopolymer solutions were mixed on ice and assembled directly on the rheometer for \SI{2}{\hour} at \SI{25}{\celsius} (Actin, Vimentin) or \SI{20}{\celsius} (K8/18).
Hybridized DX tubes were carefully placed on the rheometer and allowed to equilibrate for \SI{2}{\hour} at \SI{20}{\celsius}.
To prevent both evaporation and artifacts from interfacial elasticity, samples were surrounded with sample buffer and sealed by a cap equipped with wet sponges.
A dynamic time sweep with short measurements every \SI{60}{\second} at frequency of \SI{1}{\hertz} and a strain of \SI{5}{\percent} was used to record filament assembly and equilibration.
$G^*(\omega)$ was measured with a dynamic frequency sweep ranging from \SIrange{0.01}{80}{\hertz} at a strain of \SI{5}{\percent}. 
Fitting was performed with a self-written script in Mathematica (Wolfram Research, USA).

The differential shear modulus $K=\text{d}\sigma/\text{d}\gamma$ was obtained from transient step rate measurements at strain rates of \SI[per-mode=reciprocal]{0.025}{\per\second},\SI[per-mode=reciprocal]{0.1}{\per\second}, and \SI[per-mode=reciprocal]{0.25}{\per\second} directly after $G^*(\omega)$ measurements.
The resulting stress-strain curves were smoothed with a spline fit in \textsc{MatLab} (MathWorks, USA) and $K$ was defined as the gradient of stress $\sigma$ divided by the gradient of strain $\gamma$.

\subsection{Mesh Size}

The mesh size of a semiflexible polymer network can be estimated by assuming a simple cubic network of rigid rods with the mass per length $m_L$ and the protein concentration $c$:
\begin{equation}
	\xi = \sqrt{\frac{3 m_\text{L}}{c}}.
\end{equation}
With $m_L = \SI{2.66e-11}{\gram\per\metre}$ for F-actin \cite{steven_distribution_1983}, \SI{4.40e-11}{\gram\per\metre} for DX tubes \cite{rothemund_design_2004}, \SI{5.48e-11}{\gram\per\metre} for vimentin IF \cite{herrmann_structure_1996,wickert_characterization_2005}and \SI{3.15e-11}{\gram\per\metre} for keratin K8/K18 IF \cite{herrmann_characterization_1999} the employed concentrations ($c_\text{actin} = \SI{0.5}{\gram\per\litre}$, $c_\text{DX} = \SI{1.0}{\gram\per\litre}$, $c_\text{vimentin} = \SI{1.0}{\gram\per\litre}$, $c_\text{keratin} = \SI{0.5}{\gram\per\litre}$)  should lead to networks with similar mesh sizes ($\xi_\text{actin} = \SI{0.40}{\micro\metre}$, $\xi_\text{DX} = \SI{0.36}{\micro\metre}$, $\xi_\text{vimenin} = \SI{0.41}{\micro\metre}$, $\xi_\text{keratin} = \SI{0.43}{\micro\metre}$.

\subsection{Reptation Measurements}

Samples for single filament observations were prepared and analyzed as described previously \cite{golde_composite_2018}.
Both fluorescently labeled actin and vimentin were polymerized for one hour at room temperature.
Labeled filaments were gently mixed with unlabeled monomers to a molar ratio between 1:2000 and 1:20000 and polymerized for one hour at \SI{37}{\celsius}.
($\pm$)-6-Hydroxy-2,5,7,8-tetramethylchromane-2-carboxylic acid (Trolox - Sigma-Aldrich Co., USA) was added to a final concentration of \SI{2}{\milli\Molar} as an anti-photobleaching agent due to its radical scavenging and antioxidant activities.
Labeled DX tubes were carefully pipetted into an unlabeled DX tube network containing no anti-photobleaching agents.
The mixtures of labeled filaments embedded in an unlabeled network were placed between two glass slides, as described by Golde \textit{et al.} \cite{golde_fluorescent_2013}.
F-actin samples were kept at room temperature for one hour prior to observation.
Specimen with pure vimentin were polymerized directly in the sample chamber for two hours at room temperature.
DX tube samples were left to equilibrate overnight at room temperature.

Images of the embedded tracer filaments were recorded via an epifluorescence microscope (Leica DM-IRB, 100x oil objective, NA 1.35 - Leica Camera AG, Ger) equipped with a CCD camera (Andor iXon DV887 - Andor Technology Ltd, UK).
At least 10 filaments were captured in each sample with a frame rate of \SI{10}{\hertz} for \SI{10}{\second}.
These filaments were chosen to be well away from the glass surface and had to lie within the focal plane to enable 2D tracking.
Filament tracking was performed with the freely available ImageJ plugin JFilament (http://imagej.nih.gov/ij/).

All images of a single filament were summed up and a mean tube backbone was tracked from this overlay.
For the MSD, the filament center was defined as the point at the backbone with an equal distance to both ends.
Its movement was analyzed as a projection on the tangent vector of the tube backbone at the corresponding position.
Our definition of the filament center is susceptible to fluctuations of the contour length caused by tracking errors and filament ends moving out of focus.
Thus, we compared the MSD of the filament center to the MSD of the contour length over time divided by 4.
Filaments with a non-constant MSD of the contour length were excluded from analysis.
For filaments where both the MSD of the contour length and the MSD of the filament center are constant and comparably small, the latter is only an upper bound of the actual filament movement.

\subsection{Contour Length}

The contour length of DX tubes was determined as the median length of more than 100 DX tubes absorbed on a glass surface.
The histogram of the contour length of F-actin, vimentin IF and DX tubes is presented in Fig. \ref{fig:LC}.
The contour length of keratin K8/18 IF was assumed to have the same value as vimentin IF.
This assumption is justified by the observation that keratin and vimentin IF have a very similar length distribution for longer times despite a faster initial annealing of keratin IF \cite{mucke_vitro_2016}.

\subsection{The glassy wormlike chain model}

The specific GWLC used for this study has been comprehensively described previously \cite{kroy_glassy_2007,semmrich_glass_2007, golde_composite_2018}.
In general, the GWLC is an extension of the wormlike chain (WLC) for semiflexible polymer networks that takes into account the interactions of a test chain with its environment by stretching the mode relaxation spectrum of the WLC exponentially. 
Starting with the mode relaxation times of all eigenmodes of (half-) wavelength $\lambda_n = L/n$ and mode number $n$ for a WLC with persistence length $l_\text{p}$ and the transverse drag coefficient $\zeta_{\perp}$:
\begin{equation}\label{eq:tau_WLC}
	\tau_{n}^{\text{WLC}}=\zeta_{\perp} /(l_\text{p} k_\text{B} T \pi^4 / \lambda_n^{4} + f \pi^2 / \lambda_n^{2}),
\end{equation}
the relaxation times of the GWLC are modified according to:
\begin{equation}\label{eq:GWLC_tau}
\tau_{n}^{\text{GWLC}} = 
\begin{cases} 
\tau_{n}^{\text{WLC}} \hspace{0.7cm}   \hspace{0.7cm}  &\mbox{ if } \lambda_{n} \le \Lambda \\ 
\tau_{n}^{\text{WLC}} \text{e}^{\varepsilon N_n}         &\mbox{ if } \lambda_{n} > \Lambda.  
\end{cases}
\end{equation}
Here, $N_n = \lambda_n / \Lambda - 1$ is the number of interactions per length $\lambda_n$, $L$ the contour length of the test filament, $\Lambda$ the typical distance between two interactions, and $f$ describes a homogeneous backbone tension accounting for existing pre-stress.
$\varepsilon$ is the stretching parameter controlling how strong the modes are slowed down by interactions with the environment.
The complex linear shear modulus in the high frequency regime is then: 
\begin{equation}\label{eq:GWLC_G_supp}
G^{*}(\omega) = \Lambda/(5 \xi^2 \chi(\omega)),
\end{equation}
with the mesh size $\xi$. 
$\chi(\omega)$ is the micro-rheological, linear response function to a point force at the ends of the GWLC:
\begin{equation}\label{eq:GWLC_chi}
\chi(\omega) = \frac{L^4}{\pi^4 l_\text{p}^{2} k_{\text{B}}T } \sum_{n=1}^{\infty} \frac{1}{\left(n^4 + n^2 f/f_\text{E}\right)(1+ i \omega  \tau_{n}^{\text{GWLC}}/2) }.
\end{equation}
Here, $f_\text{E} =l_\text{p} k_{\text{B}}T \pi^2 / L^2$ is the Euler buckling force.
$f$ is set to zero for the linear regime.

In the non-linear regime, the differential shear modulus $K=\text{d}\sigma/\text{d}\gamma$ is approximated via Eq.(\ref{eq:GWLC_G_supp}) at a constant frequency as a function of the backbone tension $f$:
\begin{equation}\label{eq:K-G}
K(f) = |G^{*}_{\omega}|(f) ,
\end{equation}
where $f$ is related to the macroscopic stress $\sigma$ via $f=5\sigma\xi^2$.
The effect of pre-stress on the stretching parameter is introduced via a linear barrier height reduction:
\begin{equation}\label{eq:epsilon_f}
\varepsilon \rightarrow \varepsilon - f \delta/ k_{\text{B}}T,   
\end{equation} 
where $\delta$ should be interpreted as an effective width of a free energy well.
The mean values of $\xi$, $\Lambda$ and $\varepsilon$ obtained from fitting the linear regime for each polymer type were used to replicate the measured curves.
$\delta$ was used as the only free parameter to effectively shift the peak of $K$ both in terms of $\sigma$ and the maximum value $K_{\textrm{max}}$. 

An important question is how the other parameters are related to bottom up physical properties.
The contour length $L$, for example, is naturally a broad distribution instead of a single value [FIG. \ref{fig:LC}].
Different shapes and widths of this distribution might influence the network properties in a way that cannot be captured by a single number.

The mesh size $\xi$ is rather an effective concentration scaling than the actual distance between neighboring filaments, although it has the right order of magnitude.
The pre-factor in Eq.(\ref{eq:GWLC_G_supp}) originates from a purely geometric definition of the mesh.
A quantitative matching of $\xi$ with rheological data has been proven to be difficult for both F-actin and IF \cite{morse_viscoelasticity_1998-1,pawelzyk_attractive_2014}.

The interaction length $\Lambda$ is the average contour length between two sticky interactions of a test polymer.
Thus, it is considered as a smeared out version of the entanglement length $L_e$ in the original paper by Kroy and Glaser \cite{kroy_glassy_2007} although the GWLC is fundamentally different to the picture of a coarse grained tube.
A strict identification of both appears to be too simple and the physical nature of $\Lambda$ is still a matter of debate.
In the tube model, $L_e$ has a simple scaling of the form $L_e \propto l_p^{1/5} \xi^{4/5}$ \cite{morse_viscoelasticity_1998-2} while more advanced approaches lead to slightly different exponents \cite{morse_tube_2001}.
As expected, we cannot observe a systematic scaling of $\Lambda$ with either persistence length $l_\text{p}$ or with mesh size $\xi$ [FIG. \ref{fig:ML}].
Its consistency for DX tubes and vimentin and keratin IF might contain some information about polymer specific interactions while the strong variation of $\Lambda$ for F-actin is a direct consequence of the sample to sample variation of the cross-over frequency $\omega_\Lambda$. 
The final interpretation of the interaction length $\Lambda$ remains an important task for future investigations due to its strong influence on the transition between single polymer and interaction dominated network properties.
\clearpage

\section*{Supplementary figures}

\begin{figure}[h]
	\centering
		\includegraphics[width=0.98\textwidth]{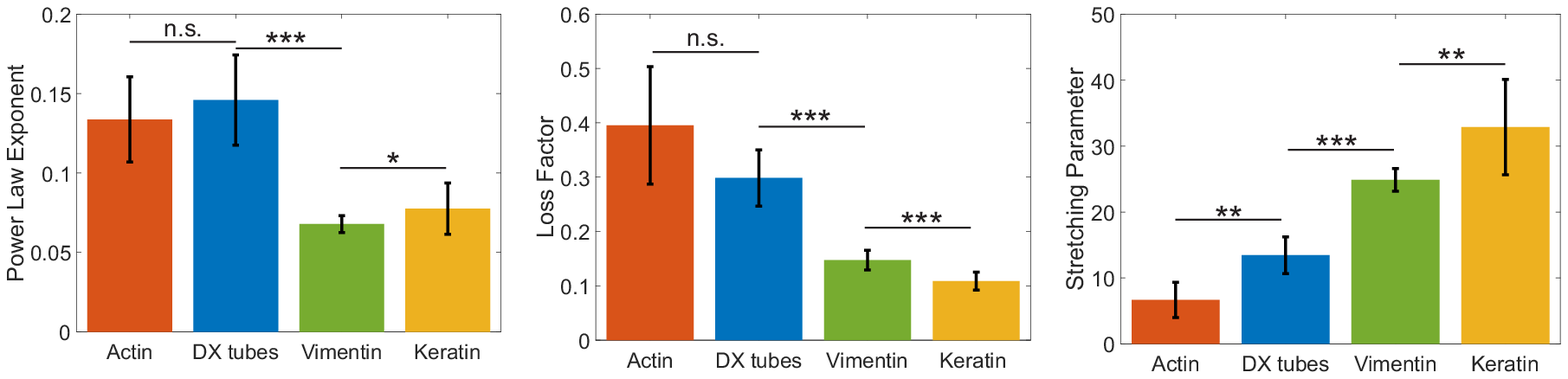}
	\caption{Local power law exponent $\alpha$, loss factor $\text{tan}(\phi)$ at a frequency $f = \SI{1}{\hertz}$, and stretching parameter $\varepsilon$ for F-actin, DX tubes, vimentin and keratin IF. Note that $\alpha$ and $\text{tan}(\phi)$ are only approximations of the actual rheological properties. Differences of  $\alpha$ and $\text{tan}(\phi)$ are not significant for F-actin and DX tubes while both polymers behave significantly different to IF. $\varepsilon$ combines both network properties and is significantly different for all four polymers.  Each bar is the mean value of all samples with $n \geq 7$. Error bars are the standard deviation of the mean. Significance was tested with a Kolmogorow-Smirnow-test. }	
	\label{fig:TanP}
\end{figure}

\begin{figure}[h]
	\centering
		\includegraphics[width=0.5\textwidth]{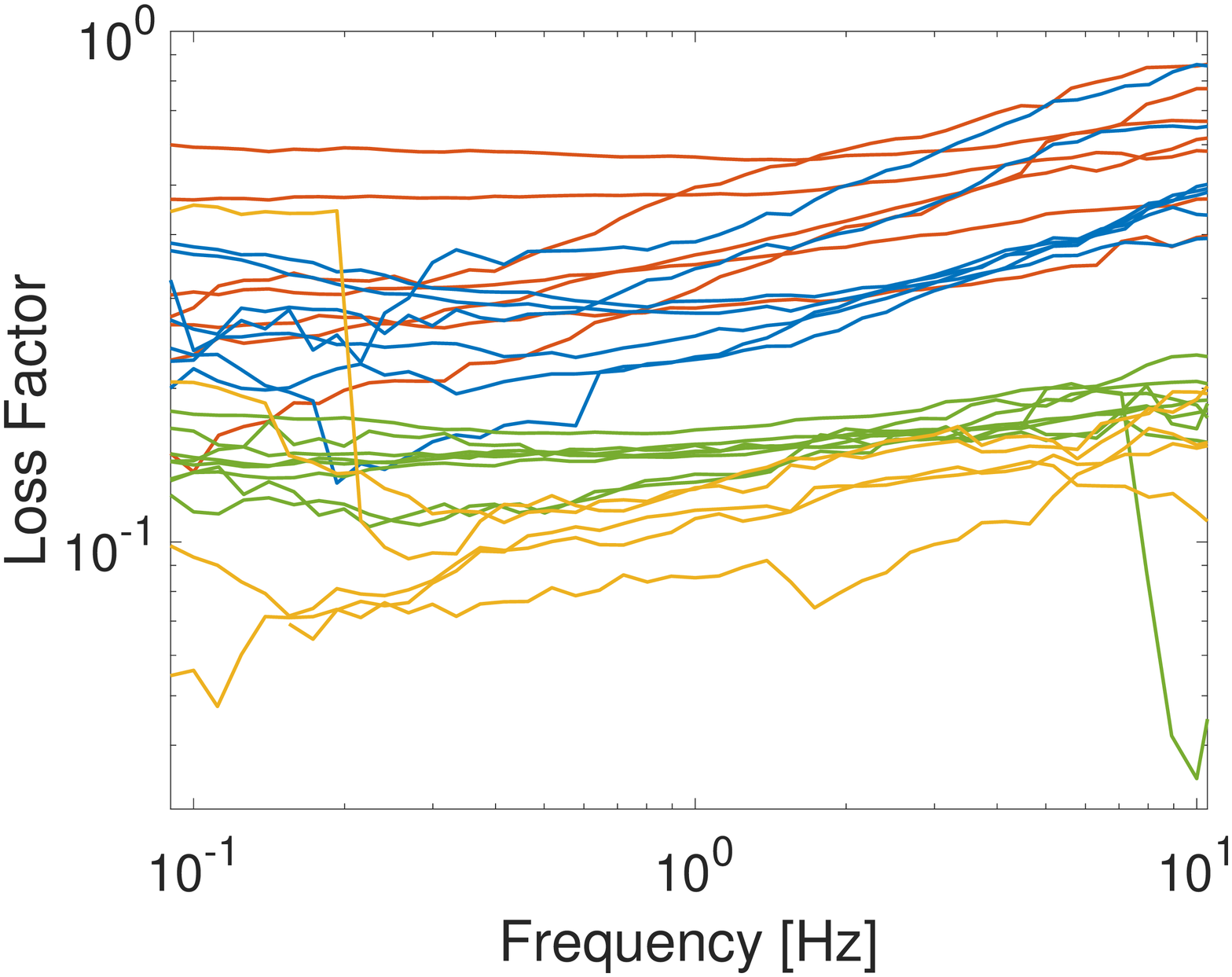}
	\caption{Loss factor $\text{tan}(\phi)$ versus frequency for F-actin (red), DX tubes (blue), vimentin (green) and keratin (yellow) IF. Each line is a single measurement. Data has been smoothed with a moving average for better visibility.}	
	\label{fig:TanP}
\end{figure}

\begin{figure}[h]
	\centering
		\includegraphics[width=0.8\textwidth]{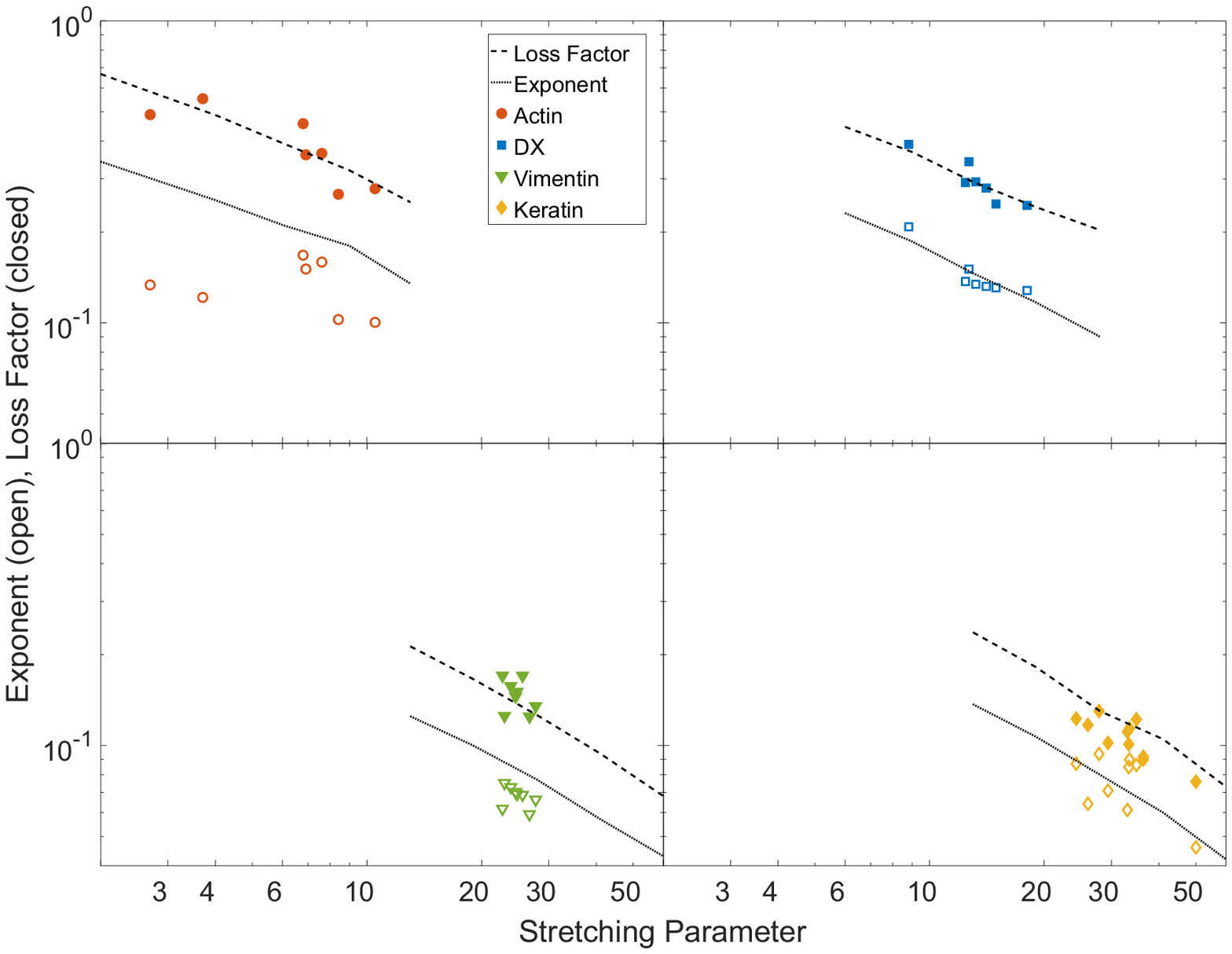}
	\caption{Local power law exponent of $G' \propto \omega^\alpha$ (open symbols) and loss factor $\textrm{tan}(\phi)=G''/G'$ (solid symbols) versus stretching parameter $\varepsilon$. Each pair of data points represents one sample. The exponent was obtained from fitting $G'$ with a power law for frequencies smaller than the cross-over between $G'$ and $G''$. The loss factor was obtained from fitting $\textrm{tan}(\phi)$ locally with a power law at a frequency of \SI{1}{\hertz}. $\varepsilon$ is the result from fitting the complex shear modulus $G^*$ to Eq.(\ref{eq:GWLC_G_supp}) for each sample. Dashed lines are the numerical results of an exemplary $G^*_\text{GWLC}$ where all parameters except $\varepsilon$ are fixed to the mean values of the polymer.}	
	\label{fig:EpsFit}
\end{figure}

\begin{figure}[h]
	\centering
		\includegraphics[width=0.5\textwidth]{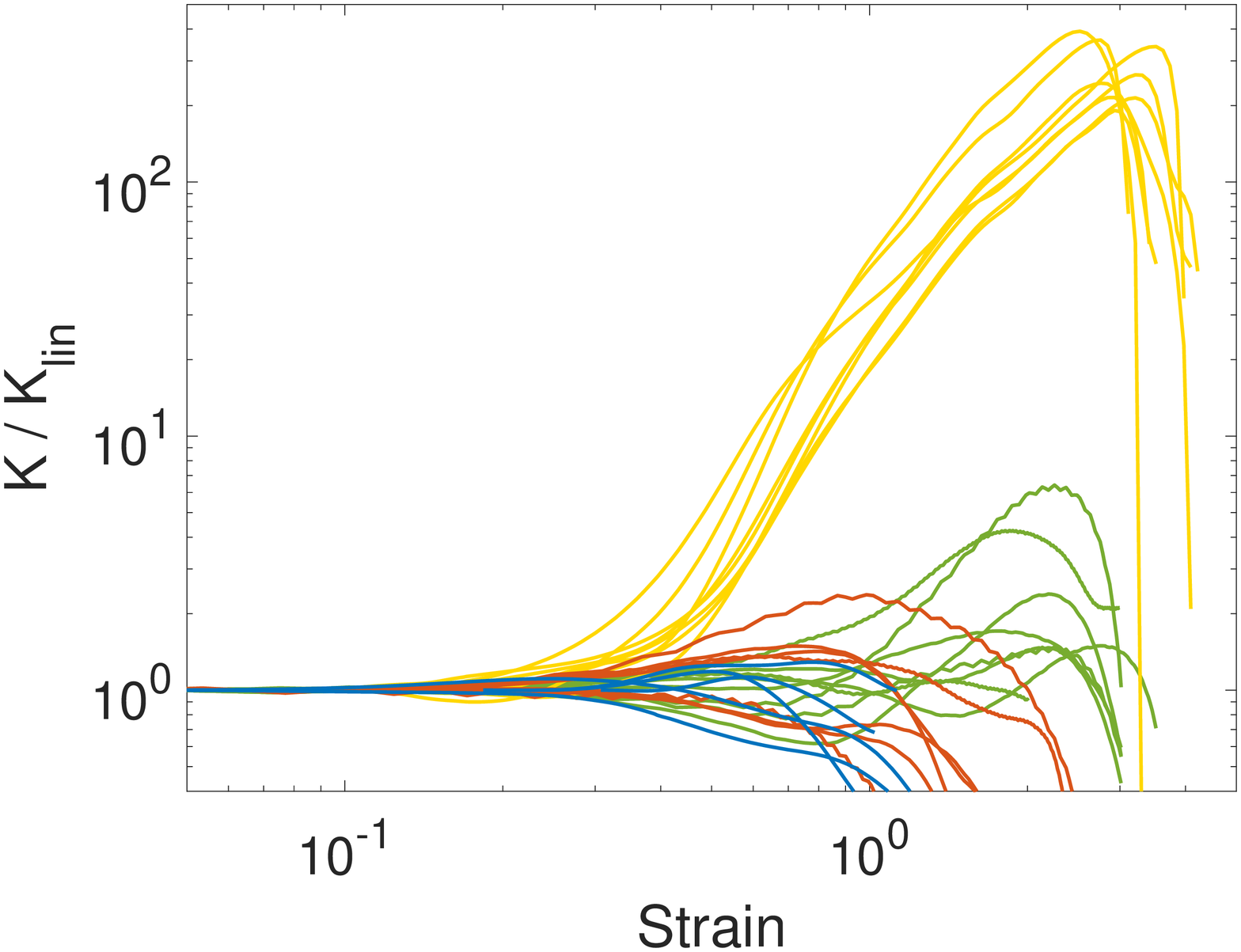}
	\caption{Differential shear modulus $K$ rescaled by its value in the linear regime $K_{\text{lin}}$ versus strain. Solid lines are single measurements of F-actin (red), DX tubes (blue) vimentin (green) and keratin (yellow) IFs samples. Actin and vimentin data reproduced from \cite{golde_composite_2018}.}	
	\label{fig:DiffK}
\end{figure}

\begin{figure*}[h]
	\centering
		\includegraphics[width=0.95\textwidth]{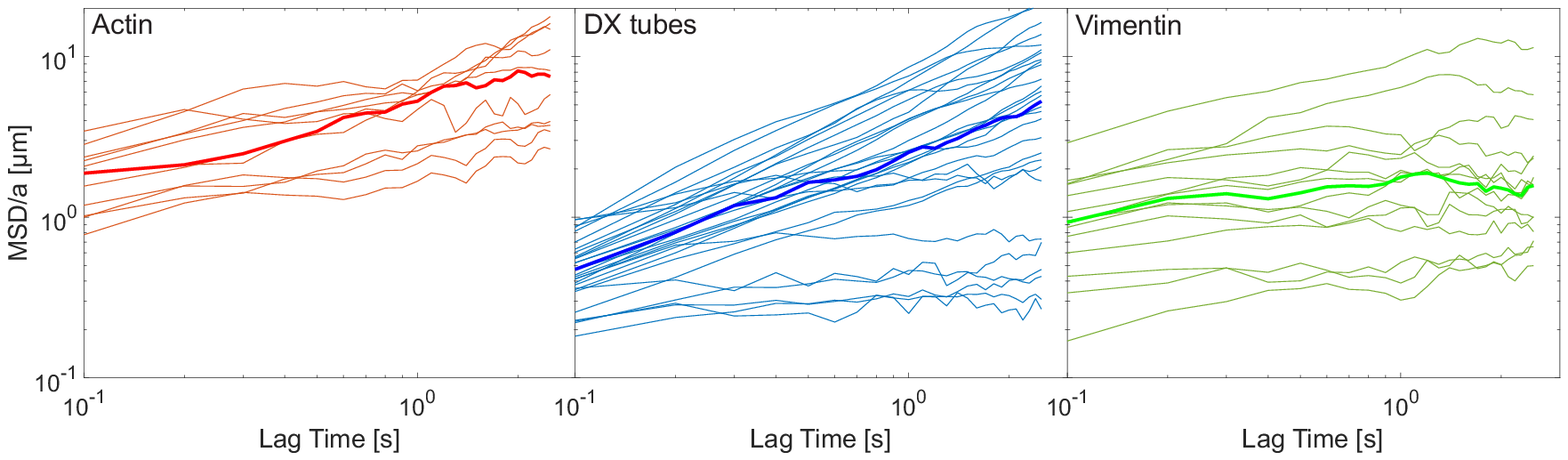}
	\caption{MSD of the filament center parallel to the tube rescaled by tube width $a$ versus lag time $\tau$. Thin lines are single filaments. Thick lines are the median over all presented filaments. Actin and vimentin data reproduced from \cite{golde_composite_2018}.}
	\label{fig:MSD}
\end{figure*}

\begin{figure}[h]
	\centering
		\includegraphics[width=0.5\textwidth]{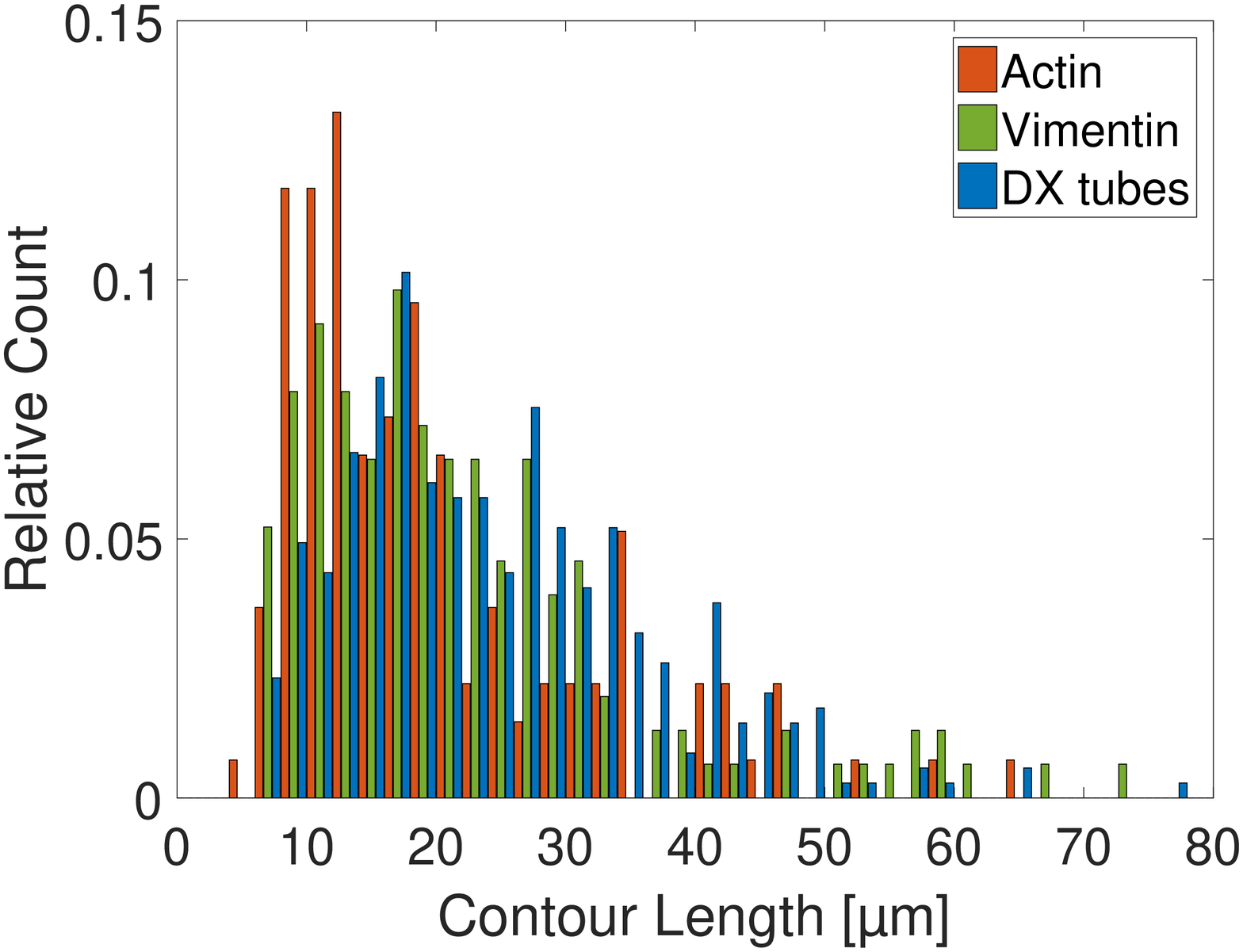}
	\caption{Histogram of the contour length $L$ of F-actin ($n=136$), vimentin IF ($n=153$) and DX tubes ($n=345$). Actin and vimentin data reproduced from \cite{golde_composite_2018}.}	
	\label{fig:LC}
\end{figure}

\begin{figure}[h]
	\centering
		\includegraphics[width=0.5\textwidth]{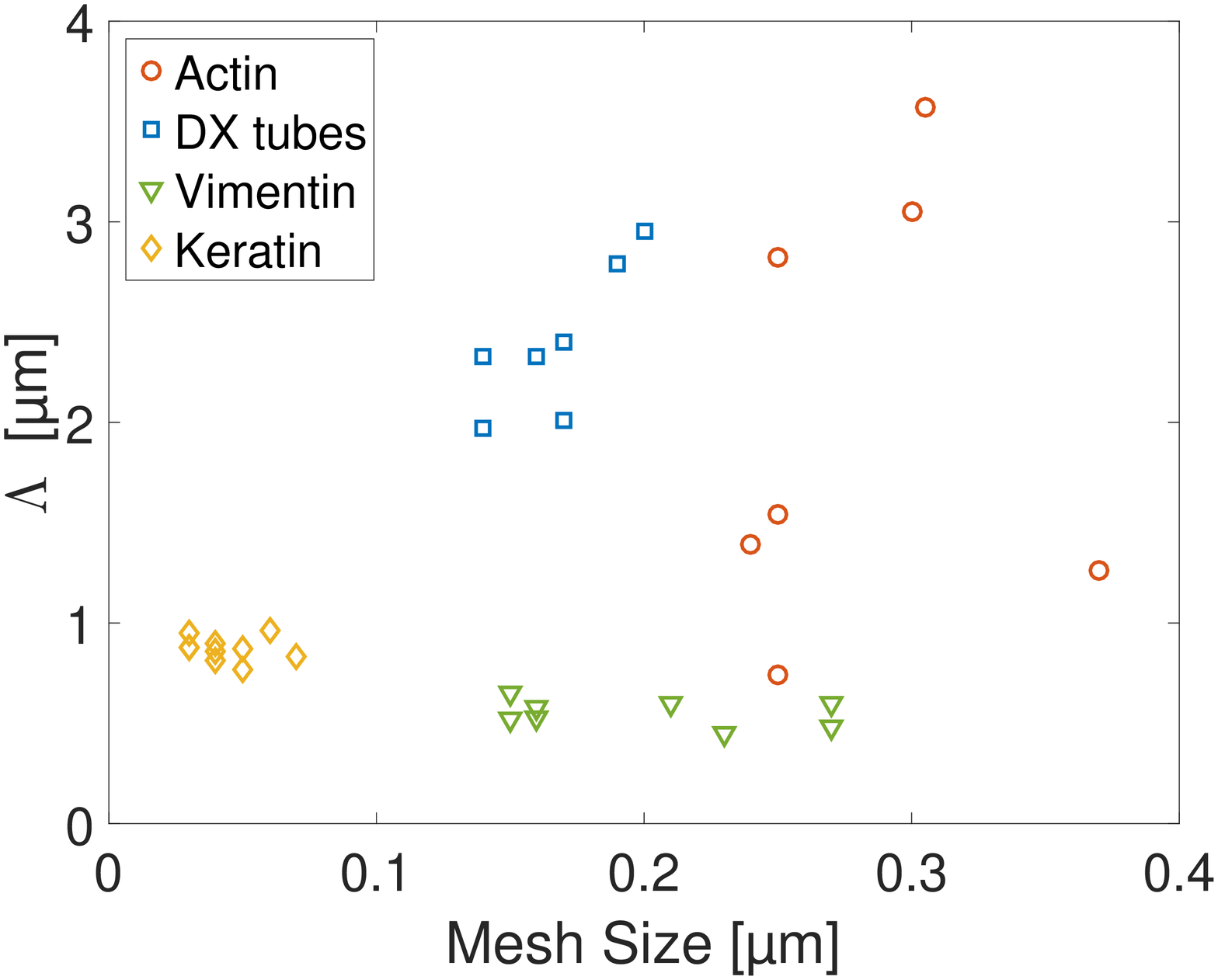}
	\caption{Interaction length $\Lambda$ versus mesh size $\xi$. All values were obtained from fitting the complex shear modulus $G^*$ of each sample to the GWLC.}	
	\label{fig:ML}
\end{figure}

\clearpage

\section*{Supplementary tables}

\begin{table*}[h]
  \centering
  \begin{tabular}{|l|c|c|}  
\hline 
linear rheology:                &                   &                       \\ 
\hline			
contour length actin                & $L$             	& \SI{16}{\micro\metre}	\cite{golde_composite_2018}\\
contour length DX tubes             & $L$               & \SI{21}{\micro\metre}	\\
contour length vimentin             & $L$               & \SI{18}{\micro\metre}	\cite{golde_composite_2018}\\
contour length keratin              & $L$               & \SI{18}{\micro\metre}	\\
contour length GWLC example         & $L$               & \SI{18}{\micro\metre}	\\
persistence length actin            & $l_\text{p}$      & \SI{9}{\micro\metre}	\cite{isambert_flexibility_1995}\\
persistence length DX tubes         & $l_\text{p}$      & \SI{4}{\micro\metre}	\cite{rothemund_design_2004}\\
persistence length vimentin         & $l_\text{p}$      & \SI{2}{\micro\metre}	\cite{mucke_assessing_2004,noding_intermediate_2012}\\
persistence length keratin          & $l_\text{p}$      & \SI{0.5}{\micro\metre} \cite{lichtenstern_complex_2012,pawelzyk_attractive_2014}\\
persistence length GWLC example     & $l_\text{p}$      & \SI{4}{\micro\metre}	\\
mesh size GWLC example              & $\xi$             & \SI{0.2}{\micro\metre} \\ 
interaction length GWLC example     & $\Lambda$         & \SI{1}{\micro\metre} \\ 
drag coefficient per length         & $\zeta_\perp$     & \SI{2}{\milli\pascal\second}\\ 
\hline 
non-linear rheology:                                         &                   &\\   
\hline			
characteristic width of a free energy well	actin            & $\delta $ 	& \SI{150}{\nano\metre}	\\
characteristic width of a free energy well	DX tubes         & $\delta $ 	& \SI{2000}{\nano\metre}	\\
characteristic width of a free energy well	vimentin         & $\delta $ 	& \SI{50}{\nano\metre}	\\
characteristic width of a free energy well	keratin          & $\delta $ 	& \SI{50}{\nano\metre}	\\
energy difference between the bound and the unbound state    & $U$          	& 2.5 $k_\text{B} T $	\\
control parameter for filament lengthening                   & $S$              & 0.13	                \\
distance between bound and unbound state	             & $\Delta x$ 	& \SI{200}{\nano\metre}	\\
\hline     
\end{tabular}
  \caption{Fixed parameters for the description of the linear rheology and adjusted parameters for reproducing the non-linear rheology.}
  \label{tab:para}
\end{table*} 

\begin{table*}[h]
  \begin{tabular}{|l|l|}\hline
   Name & Sequence \\ \hline
   SE1 			& CTCAGTGGACAGCCGTTCTGGAGCGTTGGACGAAACT \\ 
   SE2 			& GTCTGGTAGAGCACCACTGAGAGGTA \\ 
   SE3 			& CCAGAACGGCTGTGGCTAAACAGTAACCGAAGCACCAACGCT \\ 
	 SE3-Cy3	& CCAGAACGGCTGTGGCTAAACAGTAACCGAAGCACCAACGCTTT-Cy3 \\
	 SE4			& CAGACAGTTTCGTGGTCATCGTACCT \\
	 SE5			& CGATGACCTGCTTCGGTTACTGTTTAGCCTGCTCTAC \\
	\hline  
	 \end{tabular}
	\caption{Sequences of the DNA oligonucleotides.}
  \label{tab:sequence}
\end{table*} 

\section*{Discussion of filament-to-filament interactions}

F-actin has been used as the model system for entangled semiflexible polymers for decades and exhibits the smallest $\varepsilon$.
The main protein interactions are electrostatic forces due to a negative surface charge resulting in a repulsive potential shielded by ions in the buffer solution.
Larger ion concentrations lead to attractive electrostatic forces causing counterion cloud condensation.
The ion concentrations used for this study, however, are well below this transition and attractive ion effects can be ruled out \cite{janmey_polyelectrolyte_2014}.
The reason for an $\varepsilon > 1$ are most likely minor impurities and aging effects that have been shown to cause batch-to-batch variations of reconstituted F-actin networks \cite{morse_viscoelasticity_1998-1,xu_mechanical_1998}.
These batch-to-batch variations are already suggested to be a consequence of very small amounts of cross-links by Morse \cite{morse_viscoelasticity_1998-1}.
A different $\varepsilon$ for different batches is further supported by the observation, that different actin preparations lead to different power law exponents of 
$G'$ \cite{xu_mechanical_1998}.
In contrast, different rheometers used for the same preparation change the magnitude, but not the power law exponent $G'$.

DX tubes seem to be similar to F-actin in regards to their polyelectrolyte properties.
It would be very interesting to compare the effective electrostatic charge with $\varepsilon$ quantitatively.
A calculation of the effective electrostatic charge of DX tubes, however, is non-trivial due to the cross-over DNA tiles structure, and the relative dielectric constant of the medium that depends on the unknown effective ion concentration in the buffer \cite{manning_limiting_1978,manning_counterion_2007}.
The comparison of F-actin and double stranded DNA \cite{janmey_polyelectrolyte_2014} suggests a higher charge density of DX tubes, which in turn is shielded by a higher $\text{Mg}^{2+}$ concentration in the buffer.
Thus, the higher $\varepsilon$ of DX tubes is more likely a consequence of the sub-fraction of stuck filaments.
This can be explained by mishybridization during the assembly leading to strong filament connections that are independent of electrostatic interactions.

IF like vimentin and keratin filaments are known to be dominated by hydrophobic interactions \cite{yamada_mechanical_2003,pawelzyk_attractive_2014}.
These interactions are stronger for keratin due to electrostatic repulsion between vimentin IF, which partially mitigates hydrophobic attractions \cite{schopferer_desmin_2009}.
Keratin IF also express a tendency to form bundled and even clustered network structures at comparably low densities \cite{kayser_assembly_2012,deek_mechanics_2016}.
Bundling and clustering, however, should not appear at the protein concentration and buffer conditions employed \cite{leitner_properties_2012,pawelzyk_mechanics_2013}.


\section*{References}

\bibliography{fullBib}

\end{document}